\documentclass[prl,twocolumn,english,showpacs]{revtex4-1}
\usepackage[utf8]{inputenc}
\usepackage[english]{babel}
\usepackage{amsmath}
\usepackage{amsfonts}
\usepackage{amssymb}
\usepackage{lipsum}
\usepackage{graphicx}
\newcommand{\ket}[1]{\left|#1\right\rangle}
\newcommand{\bra}[1]{\left\langle#1\right|}
\newcommand{\element}[1]{${}^{#1}$}

\newcommand{\spectralLvlMath}[3]{{}^{#1}\text{#2}{}_{#3}}

\begin{document}
	
	\title{Secure distribution of a certified random quantum key using an entangled memory qubit}
	
	\author{Pascal Kobel, Ralf A. Berner and Michael K\"ohl}
	
	\affiliation{Physikalisches Institut, University of Bonn, Wegelerstra{\ss}e 8, 53115 Bonn, Germany}
	
	\begin{abstract}
		Random generation and confidential distribution of cryptographic keys are fundamental building blocks of secure communication. Using quantum states in which the transmitted quantum bit is entangled with a stationary memory quantum bit allows the secure generation and distribution of keys to be based on fundamental properties of quantum mechanics. At the same time, the reach of secure communication networks can be enhanced, in particular, since this architecture would be compatible with quantum repeaters which are an integral part for scaling quantum networks. Here, we realize a true single-photon quantum key distribution protocol (BBM92 protocol) at a second-order temporal correlation of $ {g^{(2)}(0)=0.00(5)}$ involving an entangled memory qubit which enables us to produce a certified random secret key on both endpoints of the quantum communication channel. We certify the randomness of the key using the min-entropy of the atom-photon state arising from the violation of the CHSH version of the Bell inequality of $ 2.33(6)$. 
	\end{abstract}
	
	\maketitle

\section{Introduction}

Securing communication is becoming an increasingly important task for all aspects of today's information technology. The need for data security is ranging from smart and mobile devices to global companies keeping their business secrets safe. The rise of quantum computers promises effective algorithms endangering some aspects of modern classical cryptography \cite{Shor1994}. At the same time, quantum technologies offer new ways of unconditional and proven secure communication \cite{QuantumKeyDistrBB84}. As an example, quantum key distribution (QKD) offers a paradigm-changing solution to the key distribution problem \cite{Masanes2011}. It is most commonly realized using asymmetric cryptography in classical information theory, which is considered to become insecure in the post-quantum era.   

Due to its simple prepare-and-measure architecture, the one-qubit BB84 protocol  \cite{BB84} was the first QKD protocol realized in an experimental setup. Here, the transmitted quantum bit is prepared in a certain state and basis by the sender and since the protocol does not require entangled quantum states, it was first used with weak coherent light pulses \cite{Marand:95,doi:10.1063/1.1738173}. However, the presence of pulses containing two or more photons potentially leaks information towards an eavesdropper \cite{PhysRevLett.94.230504}. This can be avoided by using true single-photon sources allowing for provably secure implementations of the BB84 protocol \cite{Waks2002}. However,  implementations of the BB84 protocol require a direct link between the communication parties. This naturally comes to a physical limit when the loss of the communication channel over distance is considered \cite{Takeoka2014} and exchanging keys over long distances requires trusted nodes as relay stations or, for ultimate security, quantum repeaters.

\begin{figure*}
	\centering
	\includegraphics[width=1\textwidth]{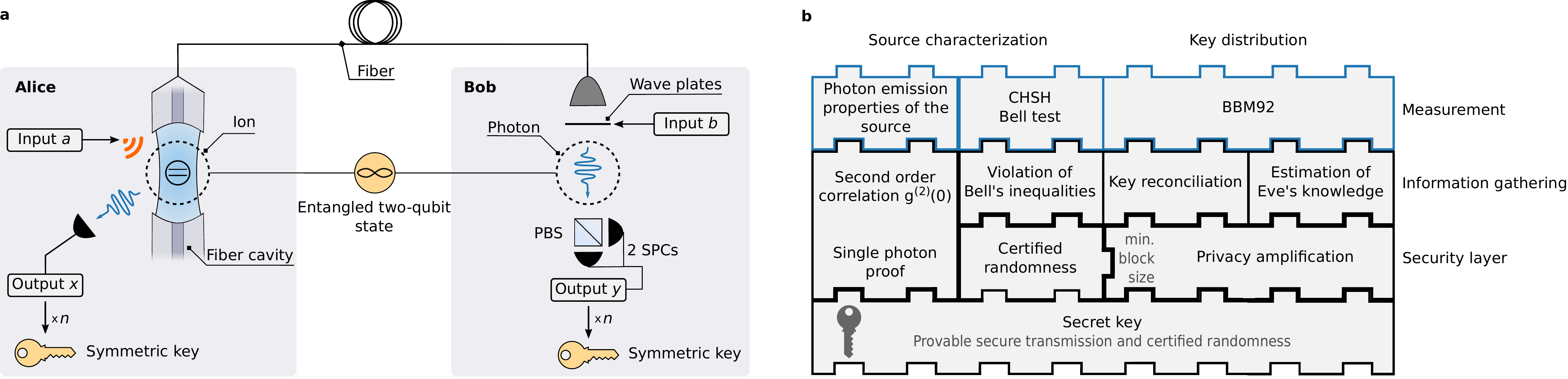}
	\caption{{\bf Experimental building blocks.} {\bf a} Sketch of the setup used for a quantum key distribution between two remote parties, Alice and Bob. The sender side (Alice) comprises a single trapped ion inside a fiber Fabry-Perot cavity. The ion emits a photon whose polarization state is entangled with the spin state of the ion. The receiver side (Bob) comprises a photon detector and adjustable polarization optics to detect the photon polarization state in different bases. {\bf b} We base the security of our key derivation on various components. Measurement blocks exhibit a blue border whereas blocks of classical data processing are sketched with black borders. } 
	\label{fig_intro}
\end{figure*}

Since a single quantum cannot be cloned \cite{NonCloningTheorem}, a QKD protocol compatible with a quantum repeater naturally has to be based on entangled-state distribution rather than preparing a specific state in a certain basis and sending it over the wire. In addition, by using stationary memory qubits  entangled with communication qubits, the fundamental point-to-point limit can be surpassed \cite{Bhaskar2020}. However, the implementation of a QKD protocol between two distant parties including an entangled memory qubit has not yet been demonstrated.

Here, we distribute a quantum key between two separate communication parties using the entanglement-based BBM92 protocol with two orthogonal bases for state projection \cite{PhysRevLett.68.557}.  We use a trapped \element{171}Yb ion embedded into a fiber Fabry-P\'{e}rot resonator as stationary qubit \cite{Steiner2013,Kobel2021}. The stationary qubit serves two purposes: i) It is a true single-photon source as the backbone for secure QKD with a second-order correlation function of $ g^2(0)=0.00(5) $. ii) It generates and stores entanglement between the spin state of the ion and the polarization state of a single emitted photon. Using the trapped ion on the sender side (Alice) and the detection of the photon on Bobs side, we securely distribute a 256-bit secret key which is usable with standard symmetric cryptographic protocols, i.e. the advanced encryption standard (AES) (see Figure \ref{fig_intro}a). It is considered that the combination of QKD and AES will provide secure communication encryption even if sufficiently large quantum computers become available that can break the classical asymmetric key distribution used today \cite{bonnetain2019quantum}.

When it comes to the generation of keys, cryptographically secure random number generators are an essential building block of secure systems. However, true random numbers are hard to generate and even harder to verify. The use of pseudo-random processes to generate secret quantities can result in pseudo-security \cite{eastlake2005randomness}. We demonstrate that we can certify the randomness of the distributed key using the fundamental non-local properties of our entangled state, which in this strong form is not possible classically. Even when using quantum systems for the distribution of keys, certifiable randomness of those quantum keys is only achieved for systems that exhibit a violation of the Bell inequalities \cite{bell2004speakable,Pironio2010}. In total, we will show that the secret key derived on both communication sides of the presented entangled system was distributed provably secure and generated with certified randomness (see Figure \ref{fig_intro}b).

\section{Experimental setup}
\subsection{ Experimental framework}
\begin{figure*}
	\centering
	\includegraphics[width=1\textwidth]{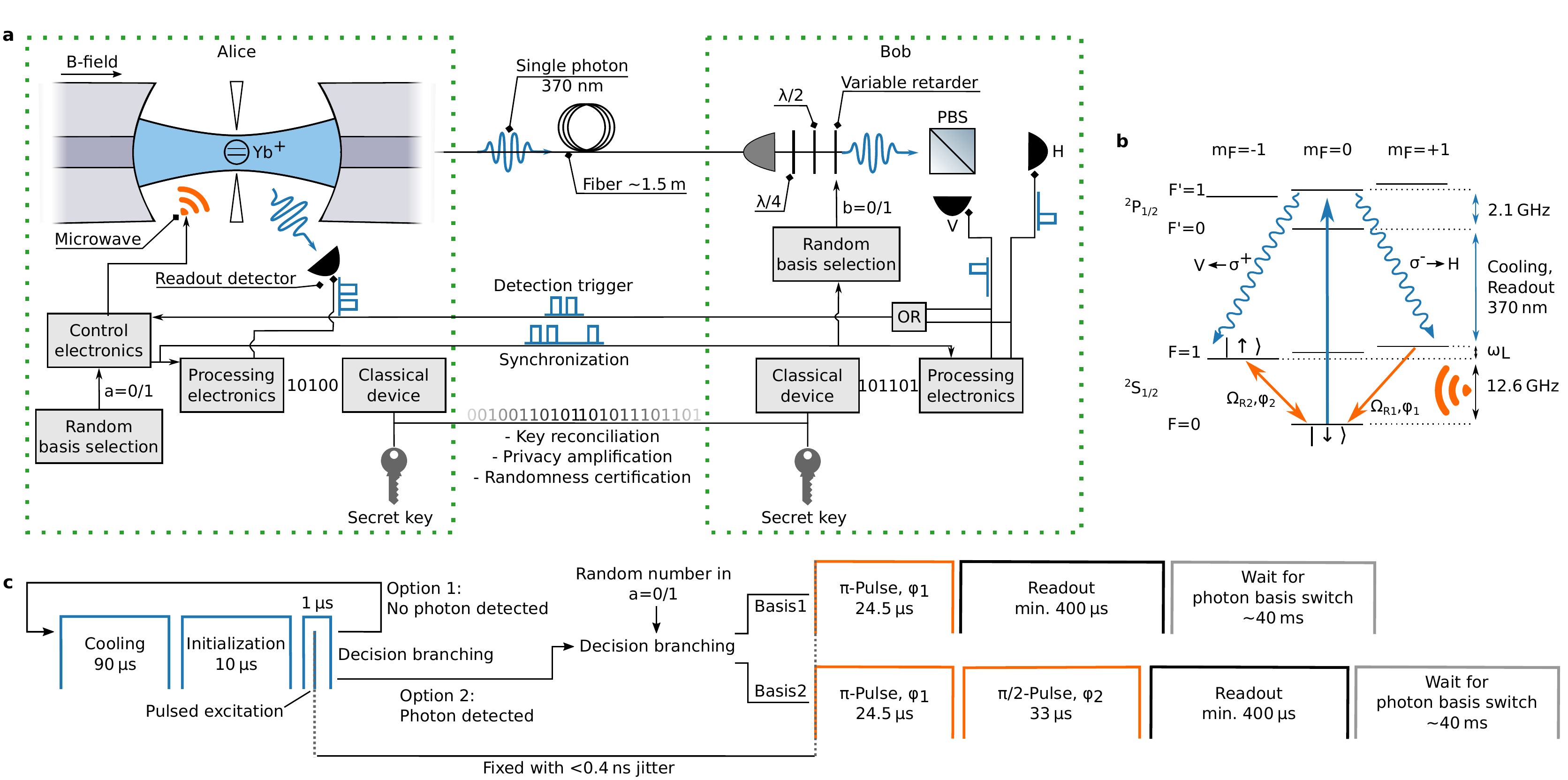}
	\caption{ {\bf Experimental realization of the QKD protocol}.  {\bf a} Experimental setup. The single emitted photons are collected along the quantisation axis on Alice's side using a fiber Fabry-Perot cavity and subsequently get transmitted to the detection side (Bob). We presuppose the existence of a public but authenticated classical communication channel between Alice and Bob for post-processing of the distributed key.
		{\bf b} Selected energy levels of \element{171}Yb\element{+} including the relevant optical transitions. After pulsed excitation from the $\ket{0}$ to the $ \ket{e} $ state, the ion decays in a superposition of decay channels emitting a $ \sigma^{\pm} $ polarized photon. Subsequent manipulation of the atomic qubit is done via microwave pulses. {\bf c} Experimental sequence on the atomic qubit side (Alice) for the measurement of one sifted key bit including the random switching of bases and the entanglement generation. Bob generates the photon detection trigger and subsequently randomly changes his detection basis, see main text for explanation.}
	\label{lvlschemeAndQubitManipulation}
\end{figure*}

Our setup comprises a single trapped \element{171}Yb ion confined in a radiofrequency Paul trap and coupled to a fiber Fabry-P\'{e}rot cavity, conceptually similar to our previous work \cite{PhysRevLett.110.043003,MSteinerMeyer,Ballance2017,Kobel2021}, see Figure \ref{lvlschemeAndQubitManipulation}a.
We initialize the ion in the hyperfine ground state $\ket{^2S_{1/2},F=0,m_F=0}$  within 4\,$\mu$s with more than $99\% $ fidelity using optical pumping with continuous-wave laser light. Here, $F$ and $m_F$ denote the hyperfine quantum numbers. We deterministically create an entangled atom-photon state $ \ket{\Psi} $ by transferring the ion to an electronically excited state on the sender side (Alice) by driving a Rabiflop using laser pulses of length $t_{\text{pulse}}=(134\pm 1 )$\,ps. The spontaneous decay of the ion within a lifetime of $ \tau=7.4(2)\,\text{ns} $ via a superposition of decay channels generates the entangled state between the ion and an emitted photon (see Figure \ref{lvlschemeAndQubitManipulation}b). The single photon is emitted by the ion into a fiber Fabry-P\'{e}rot cavity resonant to the atomic decay transition at 370\,nm. The extracted photon is intrinsically fiber-coupled, allowing easy distribution to further stages of a quantum network, i.e., a quantum repeater. Here we send the photon, which serves as a travelling qubit, via an optical fiber to a remote detection setup (Bob). 

The atomic transition linewidth of $ 2\pi\cdot\text{19.6\,MHz}$ allows for high generation rate of entanglement supported by fast extraction of photons out of the cavity of 1.3(2)\,ns. The short temporal profile of the photons is beneficial for impedance matching, i.e., to entangle with further network nodes \cite{PhysRevLett.114.123001}. 

We set the measurement basis on the atomic side of the two-qubit state using a sequence of resonant microwave pulses manipulating the spin state of the ion. Subsequently, we perform fluorescence state detection as a projective state measurement. On the photonic side the measurement basis is defined by a half- and a quarter-wave plate (HWP \& QWP) which rotate the basis of the polarization qubit. A projective measurement is achieved by a polarizing beam splitter (PBS) and two single photon counters (SPCs) on each exit path of the PBS respectively, detecting horizontal ($ H $) or vertical ($ V $) polarized photons. Active switching of the photon bases is achieved by a liquid crystal wave plate exhibiting a variable retardation of $ 0\cdot \lambda $ or $ \lambda/2 $ with the fast axis being rotated by $ 22.5^\circ $ with respect to the $ H/V $ coordinate system. Alternatively a passive basis switching would involve 4 SPCs and a 50/50 beam splitter. On the receiver side, a public authenticated channel is used for synchronisation of photon generation and detection between the two parties without revealing information about the measurement outcome or the used basis to the public.

The timing of the measurement sequence is shown in Figure \ref{lvlschemeAndQubitManipulation}c including the conditional readout of the atomic qubit which is based on the detection of the photonic qubit. Compared to previous cavity-based atom-photon entanglements, we achieved to our knowledge the yet shortest temporal shape of photons of 9.3(9)\,ns (FWHM) extracted through the cavity by more than one order of magnitude. The time profile arises from the atomic transition and cavity linewidth and is small compared to the period of the phase evolution of the entangled state ($\sim 1/5\,\mu $s) due to Larmor-precession of the atomic spin state. This allows us to project the entangled state in a single experimental shot on both sides (Alice/Bob) to a defined basis with high accuracy, which is what makes the presented QKD sequence feasible in the first place.

\subsection{ Photon state detection}
The photonic readout basis is defined by a half-wave plate, a quarter-wave plate and a variable retarder which together rotate the basis of the polarization qubit. For a defined selection of a basis, knowledge about the influence of the fiber on the polarization is required.  We characterize the photon path using a weak laser ($ \sim50 $\,pW) coupled through the PBS into the fiber.
We select the $\sigma_z$-basis by adjusting QWP and HWP to act in combination with the fiber as a quarter-wave plate mapping a circular polarized photon to a linear polarized (and vice versa) while adjusting the variable retarder to $ 0\cdot \lambda $. From this setting an orthogonal basis to $ \sigma_z $ can be selected by adjusting the variable retarder to $ \lambda/2 $. Switching the between the two retardation settings takes $ \geq 40\,\text{ms} $.

\subsection{ Atom state detection}
For a readout of the atomic state we have to map the spin states $ \ket{F=1,m_F=\pm1} $ of our 12.6\,GHz qubit to the eigenstates of the fluorescence based state detection $ \ket{F=1,m_F=-1}\equiv\ket{\uparrow}$ and $\ket{F=0,m_F=0}\equiv\ket{\downarrow}  $ using a sequence of resonant microwave pulses (see Figure \ref{lvlschemeAndQubitManipulation}b). For the $ \sigma_z $ basis we apply a $ \pi $-pulse to map $ \ket{F=1,m_F=+1} \longrightarrow \ket{\downarrow}$.
For the $ \sigma_y $ basis, the sequence of microwave pulses requires a precise timing due to the phase evolution of superposition spin states with $ 2\pi \cdot 5.477(1)\, $MHz originating from the Larmor-precession. We fix the starting time of the microwave pulses with respect to the arrival time of the excitation pulse to less than 400\,ps by synchronising them to the cavity round-trip time of the Ti:sapphire laser. 
Both pulses originate from an arbitrary waveform generator and are mixed to a carrier signal red detuned by $ \sim 8\,\text{MHz} $ from the center of the two microwave transition frequencies. Mixing to the same carrier preserves the relative phase $ \Delta \phi $ between the pulses. Since the experimental sequence is not synchronised to the microwave carrier phase, the first $ \pi $-pulse starts with a random phase with respect to the Larmor-precession of the atomic qubit. The following $ \pi/2 $ pulse acting on the $ \ket{\downarrow} $/$ \ket{\uparrow} $ qubit rotates around an axis with a fixed relative orientation to the phase of the $ \ket{\downarrow} $/$\ket{\uparrow} $ superposition. The relative orientation of this rotation axis is determined by the phase difference $ \Delta \phi $. In total, both pulses rotate the ion qubit around a fixed axis regardless of the phase of the first pulse. We end up in total with a defined atomic basis for readout by considering a fixed timing of the pulses in the laboratory frame with respect to the Larmor-precession. 
Using this technique we are able to precisely select any basis orthogonal to $ \sigma_z $ for the atomic qubit via the phase difference $ \Delta \phi $. 
\newline

\section{Results}
We present the implementation of a measurement protocol that allows two remote communication partners to exploit fundamental quantum mechanical properties such as non-locality and non-cloning of an entangled quantum state distributed between them to derive a secret key with excellent cryptographic properties. In Figure 1b, we show how the building blocks of our protocol work as a whole to ensure the crucial properties of a secret key, namely randomness, confidentiality and integrity. In the following, we will discuss these blocks in detail.

\subsection{ Key distribution protocol}
We utilise a maximally-entangled two-qubit state between a single trapped ion and a single photon \cite{Kobel2021} for distribution of a quantum key between two remote parties A and B. 
On the atomic side, the information is encoded in the electronic ground state $\spectralLvlMath{2}{S}{1/2}$ of the Yb$^+$ ion in the $\ket{F=1,m_F=-1}\equiv \ket{\uparrow} $ and $\ket{F=0,m_F=0} \equiv \ket{\downarrow} $ states. On the photonic side we employ the polarization modes $ \ket{H}/\ket{V} $ to encode information. A precise knowledge about the entangled two-qubit state is not required for the BBM92 protocol, however, we know the state from a full quantum state tomography \cite{Kobel2021} to be:
\begin{equation}
\begin{aligned}
\ket{\Psi}&=\frac{1}{\sqrt{2}}\left(\ket{V}\ket{\uparrow}-\ket{H}\ket{\downarrow}\right)\\
&\equiv\frac{1}{\sqrt{2}}\left(\ket{0}_\text{photon}\ket{0}_\text{atom}-\ket{1}_\text{photon}\ket{1}_\text{atom}\right)
\end{aligned}
\label{Eq_Zustand}
\end{equation}
with a fidelity of $ F= (90.1\pm1.7)\text{ \%} $.

We apply the entanglement-based BBM92 protocol \cite{PhysRevLett.68.557} to the two-qubit quantum state of equation (\ref{Eq_Zustand}) in order to derive a secret key on  both sides A and B, respectively, where the matter qubit remains at side A (Alice) and the photon is transmitted to side B (Bob). Alice and Bob are using random inputs $ a,b \in \{0,1\} $ to set the basis of the projective state measurement on their respective side of the two-qubit system. Alice obtains a value $ x\in \{0,1\}$ from the measurement while Bob obtains $ y\in \{0,1\}$. We choose the $ \sigma_z $ basis for projective state measurement when $ a,b=0 $ and the orthogonal $ \sigma_y $ basis when $ a,b=1 $. 

When Alice and Bob happen to measure in the same basis $ a=b $ then their outputs are equal ($ x=y $) in the ideal case as a consequence of the two qubits sharing the entangled state $\ket{ \Psi} $. However, in reality the outputs can differ occasionally due to measurement or state preparation imperfections, or due to an attacker Eve performing measurements on the quantum state. The outputs $ x_i $ and $ y_i $ of the state measurements obtained in round $ i $ when $ a_i=b_i $ form the sifted key strings $ {\bf X}=x_1,...,x_n  $ and $ {\bf Y}=y_1,...,y_n $. After a measure-and-estimate phase of the quantum bit error rate (QBER) on the sifted key strings both parties agree on continuing with key distribution in case of an acceptable QBER ($ <15\,\% $). Up to now, the sequence is quite similar to the implementation of a BB84 protocol with the exception of not preparing a certain quantum state~$ x $ in basis~$ a $ at the senders side and sending it to the receiver measuring $ y $. Instead, entanglement is distributed between the parties to make the qubits on sender and receiver side sharing a common wavefunction, which upon a state measurement gets projected to a certain basis.
In this context we will show that, in contrast to implementations of the BB84 protocol, the value of the outputs $ x,y $ are truly random.

\subsection{ Certified randomness}
Since the BBM92 and the BB84 protocol do not differ in their measurement implementation and observables, there is no way for the communicating parties to find out from the key distribution itself whether they have used entangled two-qubit states or simply used a prepare-and-measure scheme (where the prepare part was taken from the source on Alice's side). However, the randomness of the derived key depends crucially on whether the transmitted photons were part of an entangled two-qubit state or just part of a mixed state. In consequence, the communication parties need a method to unambiguously prove the generation of an entangled two-qubit state as a prerequisite for random bit generation and a mathematical expression to estimate the quantity of random bits generated. Both can be elegantly achieved in one step via the Bell inequalities.
 
First, we have to quantify the concept of randomness. We consider the uncertainty of an attacker with side information E about the system S, which can be expressed through the probability of guessing a measurement outcome~$ x $ of measurement~$ s $ on S \cite{5208530}. In the extreme case of S being fully correlated to parts of E, the guessing probability becomes $ P_\text{guess} (x)=1 $. In other cases, the state S is (partly) independent of the attackers information E and the guessing probability can be described as \cite{Masanes2011}
\begin{equation}
P_\text{guess} (x) = \max_x P(x|s)
\end{equation}
where $ P(x|s) $ is the probability of measuring output value $ x \in \{0,1\} $ for a measurement $ s$ and the maximum is taken over all possible output values $ x $. Here, $ s \in \{0,1\} $ is the respective choice of the measurement basis. 
True randomness is achieved for a sifted key string $ {\bf X}=x_1,...,x_n $ when $  P(x|s)=\text{const}, \, \forall x,s$ where all possible combinations of $ \bf {X} $  are equal probable as an outcome.

In classical information theory, the generation and even more the verification of random numbers (random bits) is hard to realize because one has to exclude any causal connections between the numbers, in particular the specification of an upper bound for $ P_\text{guess}(x) $ is a problem. In contrast, in quantum theory the unpredictability of a measurement outcome is closely linked to the violation of the Bell inequalities. Using violation of the Bell inequalities one can quantify the closeness to a situation where a quantum system S is fully determined by the side information E \cite{5208530} and when it is perfectly non-local and non-deterministic on the opposite without knowing the system's internal behaviour. Any system maximally violating the Bell inequalities exhibits a non-local, non-deterministic description, which excludes any deterministic (causal) connections and generates certified private randomness \cite{ColbeckRandomness,Pironio2010}.

We consider to test the violation of Bell inequalities in the experiment in the form proposed by Clauser, Horne, Shimony and Holt (CHSH) \cite{PhysRevLett.23.880}, where a system that can be described as local and deterministic satisfy:
\begin{equation}
g=\sum_{a,b} (-1)^{ab} \left[P_{x=y}(a,b)-P_{x\neq y}(a,b)\right] \leq 2, 
\label{Eq_CHSHInequality}
\end{equation}
where $ P_{x=y}(a,b) $ is the probability of measuring the same output on both sides when using the measurement basis $ a,b $. The quantum theory predicts a maximal violation of $ 2\sqrt{2} $.

We have measured a violation of Bell's inequality of~$ g_\text{meas}=2.33(6)$ by using the measurement basis $ \sigma_y $ for $ b=0 $ and $ \sigma_x $ for $ b=1 $ on Bobs side. On Alice side we measure at $ (\sigma_y-\sigma_x)/\sqrt{2} $ for $ a=0 $ and at $ (\sigma_y +\sigma_x)/\sqrt{2} $ for $ a=1 $. The observed outcomes of the measurements are shown in Table \ref{Table_ComparisonValuesOtherAtomPhotonEntanglement}. 
Due to the entanglement generation and detection scheme, we measure a random subset of all generated two-qubit states. For the violation of equation (\ref{Eq_CHSHInequality}), we assume a fair sampling of this subset from the total set of generated entangled states. 

We cross check the measured violation of Bell's inequality $ g_\text{meas} $ with the Bell-violation $ g_\text{exp} $ we would expect from the measured state fidelity  $ F=(90.1\pm1.7) $. For this purpose, we assume our state to be a pure state $ \ket{\psi} $ mixed to a probability $ (1-V) $ with white noise: $\rho=V\ket{\psi}\bra{\psi}+(1-V)\frac{\mathbb{I}}{4}$
where $ \mathbb{I} $ is the $ 4 \times 4 $ identity matrix.
With the visibility $ V=2F-1=0.80(3)$ we obtain $ g_\text{exp}=2\sqrt{2}\cdot V=2.27\pm 0.10 $ according to \cite{Masanes2011}, which is consistent with the measured Bell-violation $ g_\text{meas}=  2.33(6)$.

\begin{table*}[!htb]

	\begin{tabular}{| l  c c c c|}
		\hline
		$ (a,b) $ & $ (1,0) $ & $ (0,0) $ & $ (1,1)  $ &$ (0,1) $\\
		\hline
		Basis $\sigma_\text{atom} \otimes  \sigma_\text{photon}$ & $ \frac{\sigma_y+\sigma_x}{\sqrt{2}} \otimes \sigma_y$ &  $\frac{\sigma_y-\sigma_x}{\sqrt{2}} \otimes \sigma_y$ & $\frac{\sigma_y+\sigma_x}{\sqrt{2}} \otimes \sigma_x$ & $\frac{\sigma_y-\sigma_x}{\sqrt{2}} \otimes \sigma_x $ \\
		\hline
		$ P_{x=y} $& $0.835\pm0.020 $ &$ 0.791\pm0.027 $ &$ 0.229\pm0.013 $ &$ 0.770\pm0.027 $\\
		
		\hline
	\end{tabular}
	\caption{{\bf Measurement of the Bell-violation}. Observed outcome of the measurement outputs (x,y) for the binary choices of measurement bases (a,b).}
	\label{Table_ComparisonValuesOtherAtomPhotonEntanglement}
\end{table*}

  \begin{figure*}
	\centering
	\includegraphics[width=0.95\textwidth]{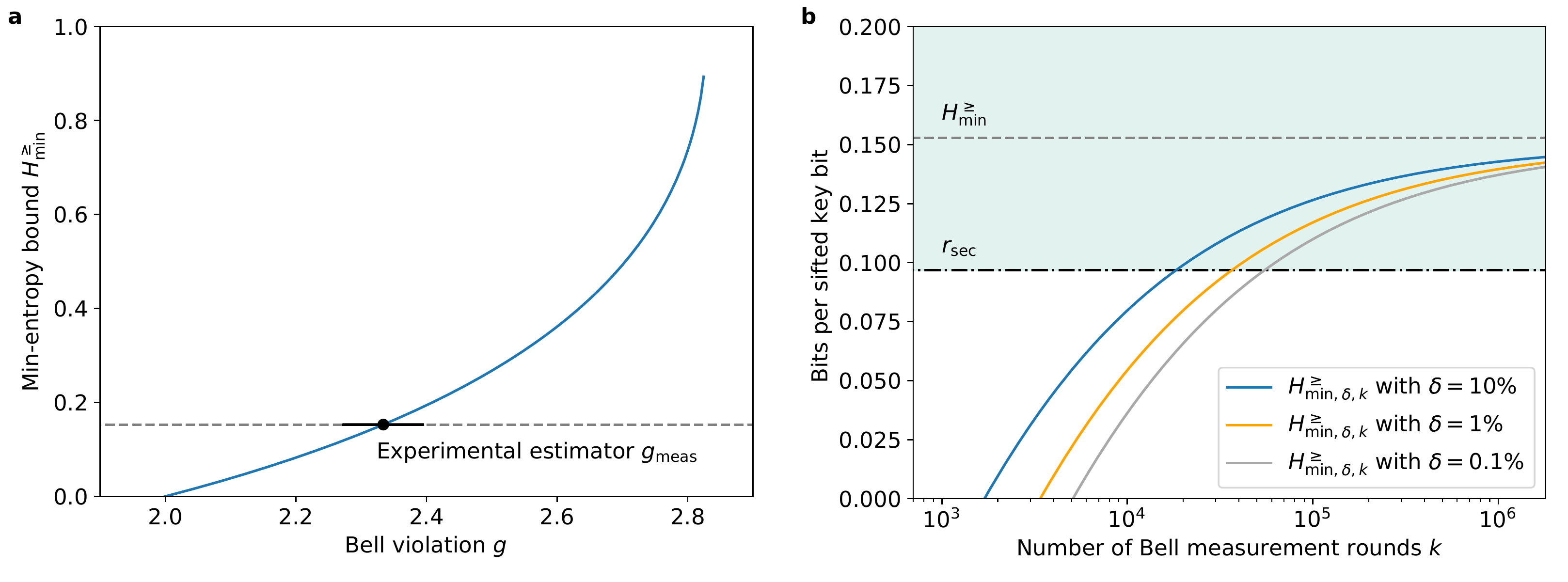}
	\caption{{\bf Generated randomness vs. Bell-violation. } {\bf a} The lower bound on the min-entropy $ H_\text{min}^{\geq} $ generated per measurement on the two-qubit state $ \ket{\Psi} $ is shown versus violation of the CHSH version of the Bell inequality $ g $ (see equation (\ref{Eq_minentropyperBitBound})). The error on the measured estimator  $ g_\text{meas} $ is the standard error. {\bf b} Lower bound on the min-entropy $ H_{\text{min},\delta,k}^{\geq} $ as a function of the number $ k $ of measurements made for the estimation of the Bell violation  $ g_\text{meas} $ with the confidence level $ \delta $ according to equation~(\ref{Eq_minentropyperBitBound_finite}) (solid lines). The grey dashed line depicts the asymptotic limit of the min-entropy bound as shown in {\bf a}. The coloured area depicts the region where the communication parties can be sure with a confidence level  $ \delta $ that the generated randomness per sifted key bit is larger than the secret key derived from the sifted key even in the asymptotic limit of a large sifted key length $ n $ (for the definition of  $ r_\text{sec} $, see equation (\ref{Eq_AsymptoticSecretKeyRate})).} 
	\label{fig_randomBits}
\end{figure*}

From the violation of the CHSH inequality $ g $ of equation~(\ref{Eq_CHSHInequality}), the upper bound on the guessing probability of each measurement outcome $ x $ by an attacker in the asymptotic limit can be derived as \cite{Pironio2010}
\begin{equation}
P_\text{guess}(x)\leq 0.5+0.5\sqrt{2-\frac{g^2}{4}}.
\label{Eq_lowerBoundOnGuessingProb}
\end{equation}
We assume that any reduction in non-locality results in an increasing causality which may be fully accessible to an attacker. For finite measurement rounds, the measured violation $ g_\text{meas} $ is an estimator of the CHSH inequality violation $ g $ and the guessing probability may be higher. It was shown in \cite{Pironio2010,PhysRevA.87.012336} that a lower bound on $ g $ can computed from the estimator $ g_\text{meas} $ as
\begin{equation}
 g\geq  g_\text{meas}-\epsilon(k,\delta)
\end{equation} when the device was used $ k$-times in succession. Here $ \epsilon(k,\delta)=\sqrt{-\frac{\ln\left(\delta\right)\cdot 2\left(1/q+g_\text{meas}\right)^2}{k}} $ with an uncertainty parameter $ \delta $ and $ q=0.25=\min_{a,b}\left[P(a,b)\right]$  being the minimum of the probability distribution over the input states $ a$ and $b $. 

For a single measurement $ s $ on the system S the min-entropy $ H_\text{min} $ quantifies the amount of randomness generated in this measurement.
It was shown, that the min-entropy per transmitted bit can be computed as \cite{5208530}
\begin{equation}
H_\text{min}(s|E)=-\log_2\left[P_\text{guess}(x)\right]
\label{Eq_minentropy}
 \end{equation}
where $ P_\text{guess}(x) $ is the probability of guessing the correct output x of measurement $ s $ by the attacker measuring the side information E \cite{Masanes2011}.

Using equation (\ref{Eq_lowerBoundOnGuessingProb}) in combination with equation (\ref{Eq_minentropy}), we compute a lower bound on the min-entropy generated per measurement on the quantum system state $ \ket{\Psi} $ as
\begin{equation}
H_\text{min}(g)\geq -\log_2\left[0.5+0.5\sqrt{2-\frac{g^2}{4}}\right] \equiv H_{\text{min}}^{\geq}
\label{Eq_minentropyperBitBound}
\end{equation}
in the asymptotic limit. Finite key sizes $ n $ can be considered using \cite{Pironio2010} 
\begin{equation}
H_{\text{min}}\left(g_\text{meas}-\epsilon(k,\delta)\right)\equiv H_{{\text{min},\delta,k}}^{\geq}
\label{Eq_minentropyperBitBound_finite}
\end{equation} 
with $ \delta $ being the confidence level of randomness generation.  
Equation (\ref{Eq_minentropyperBitBound}) quantifies the asymptotic lower bound on the generated randomness per measurement to be $ H_{\text{min}}^{\geq}=0.15(4)  $\,bits.
In Figure \ref{fig_randomBits}a the lower bound on the min-entropy $ H_{\text{min}}^{\geq} $ is shown as a function of the Bell violation $ g $. Figure \ref{fig_randomBits}b shows the subsequent lower bound on the min-entropy $ H_{\text{min},\delta,k}^{\geq} $ as a function of the number $ k $ of measurements made for the estimation $ g_\text{meas} $ of the Bell violation and the confidence level $ \delta $ according to equation~(\ref{Eq_minentropyperBitBound_finite}). 
Further, we can calculate the minimum block size $ l_B= \text{ceil} (1/H_{\text{min}}^{\geq}) =(7\pm2) \text{\,bits}$ of the sifted key for which Alice and Bob have generated at least one random bit in the asymptotic limit. The key length $ n_\text{rand} $ for a final key to be entire random is bounded by $ n_\text{rand}\leq \text{floor}( n_r/ l_B )$ with $ n_r\leq n $.

\subsection{\bf Key reconciliation}

\begin{figure}
	\centering
	\includegraphics[width=0.95\columnwidth]{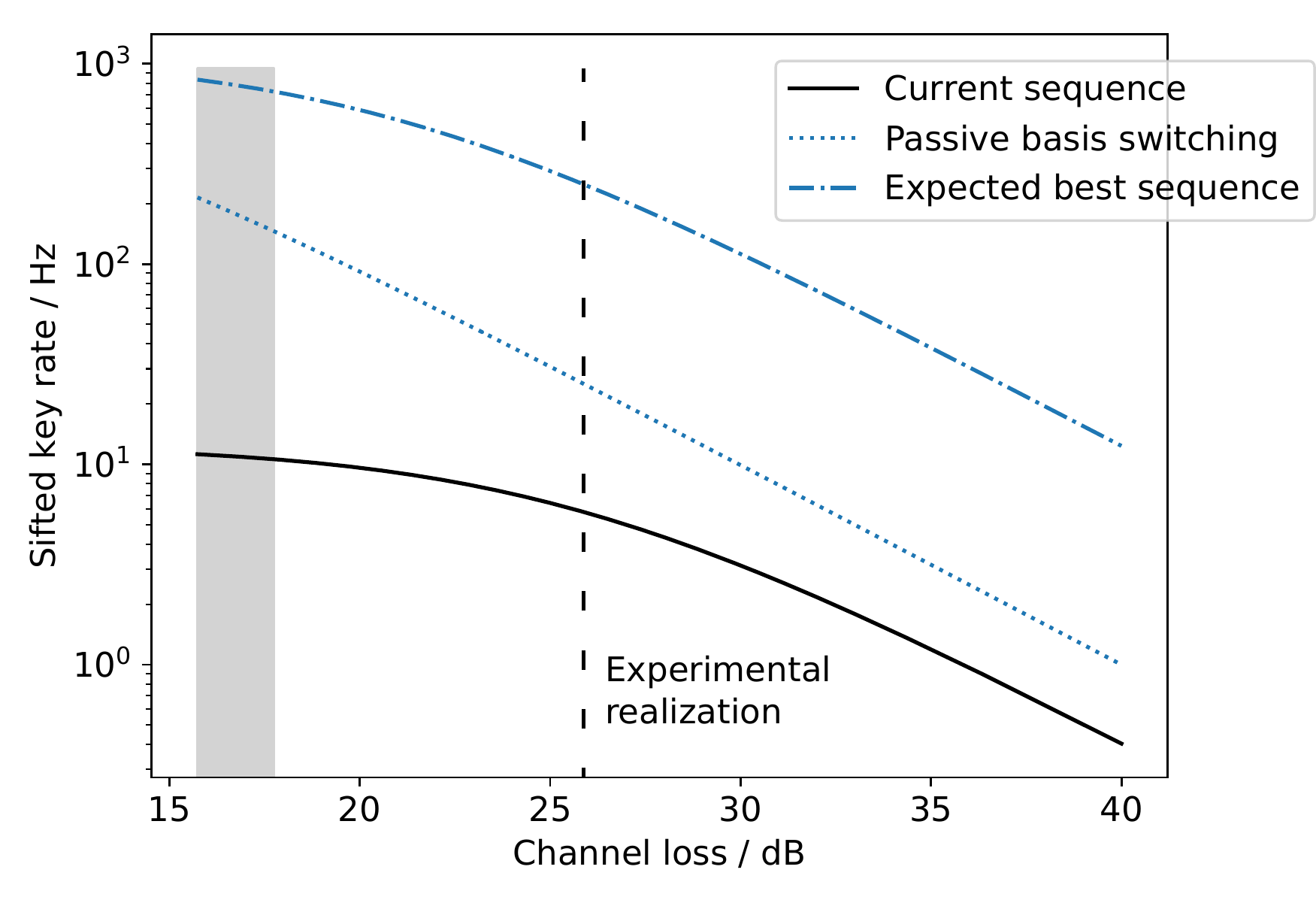}
	\caption{{\bf Calculated maximum sifted key rate vs. channel loss.} The performance of the current experimental sequence including the active switching of photon bases is shown as solid black line. The channel loss of the setup ($ 26 $\,dB) is shown as vertical dashed line. The performance of an alternative setup without active switching and using 4 single photon counters instead is shown as blue dotted line. We reached this efficiency in \cite{Kobel2021}. An estimate of a sequence where also the generation rate of entangled photons is optimised by minimising the stationary qubit initialization time to $ 4\,\mu\text{s} $ is shown as blue dash-dotted line. The theoretical reachable minimal loss of the setup of $ 16.7(10)\, $dB is shown as grey shaded area.} 
	\label{fig_siftedKeyRates}
\end{figure}

From each projective state measurement, Alice and Bob obtain one bit of their sifted keys $ {\bf X} $ and $ {\bf Y} $, each with a total length $ n $. We obtain a maximal sifted key bit rate of $ 6 $\,Hz limited by the switching time of the photon readout basis ($ 40 $\,ms), stationary qubit initialization ($ \sim 100\,\mu\text{s} $) and the channel loss (see Figure \ref{fig_siftedKeyRates}). Using a setup with passive basis switching could improve the sifted key rate by one order of magnitude. We observed this rate in terms of the two-qubit state detection with the same light-matter interface without an active switching of bases \cite{Kobel2021}. A well known passive scheme can be realised with 4 single-photon counters for state detection, with each pair of detectors separated by a 50/50 beam splitter. Further, another order of magnitude in the sifted key rate can be gained by exploiting the minimum state initialization time of $ \sim4\,\mu\text{s} $ of the memory qubit in advance to the deterministic photon generation. Currently, we always perform Doppler cooling of the ion before state initialization, which also could be done in separate cooling sequences. However, each of the presented alternatives is limited by the state detection of the ion which lasts $ \sim 400\,\mu $s.

We currently perform our experiment at a channel loss of -26 dB including path loss (-2.3(8)\,dB) and detector efficiencies (-6.7 dB). The channel loss is mainly limited by the photon extraction probability from the fibre-cavity ($ \sim $ -7.8 dB), the mechanical stability of the fiber cavity and the localization of the ion within the resonator mode (together $ \sim $-17 dB). In the future, the latter figure could be significantly improved by an revised mechanical design of the experimental setup and better ion localization.
This corresponds to a sifted key bit rate of $ 1.29(3) \times 10^{-3} $ bits per channel use, where we defined one entanglement generation attempt as a channel use.   

\begin{figure*}
 	\centering
 	\includegraphics[width=0.95\textwidth]{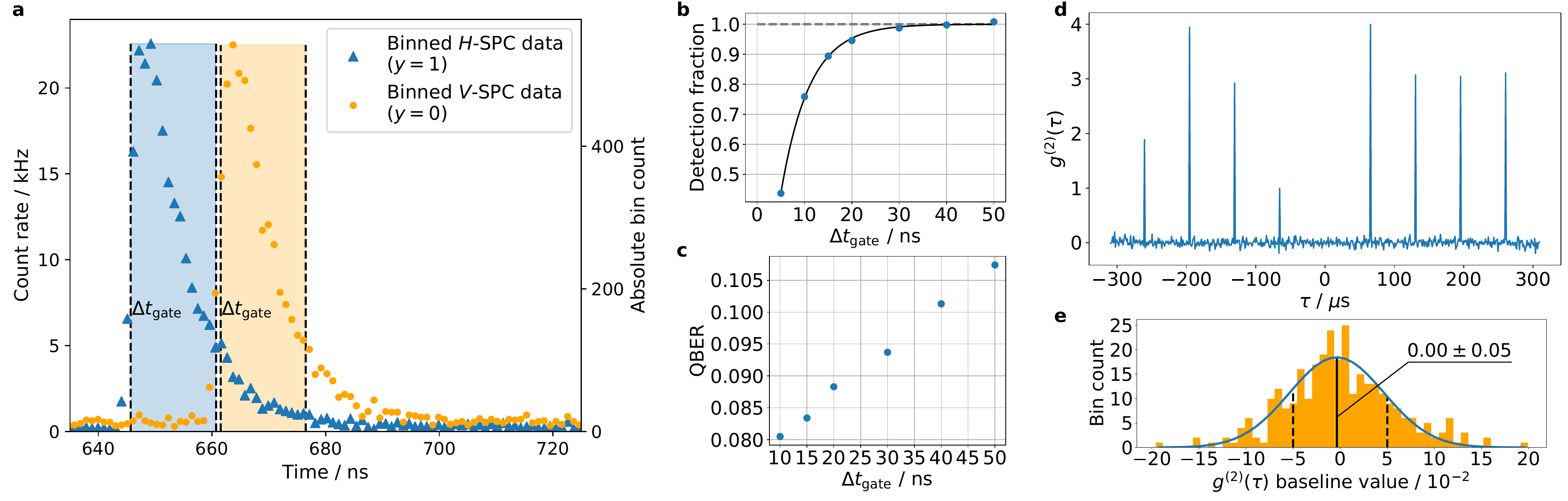}
 	\caption{{\bf Photon detection statistics.}  {\bf a} Binned photon arrival times on Bob's side.  In order to suppress dark counts, a measured bit only contributes to the sifted key for the detection to happen in the time window $ \Delta t_\text{gate} =\text{15\,ns}$ which is shown as colored  areas for $ H $/$ V $ respectively. {\bf b} The relative fraction of flying qubits contributing to the sifted key is shown for different acceptance windows (blue points). The solid line constitutes an exponential saturation fit. {\bf c} Quantum bit error rate on the whole sifted key for different acceptance windows. {\bf d} Second-order correlation function $ g^{(2)} $ of the photon arrival times measured in a HBT setup where we do the noise level correction of both detectors according to \cite{Keller2004}. We use a photon generation sequence with 17\,kHz repetition rate for the determination of $ g^{(2)}(\tau) $.  {\bf e} Binned values of the correlation baseline where we extract the mean (black solid line) and the standard deviation (black dashed lines) of the Gaussian noise (blue curve as a fit) and finally obtain $ g^{(2)}(0)=0.00(5)$.}
 	\label{fig_photonStatistic}
 \end{figure*}

The sifted keys differ in $ m $ bits due to measurement or state preparation imperfections, or due to an attacker Eve performing measurements on the quantum state. To reduce measurement errors on Bob's side, we suppress dark counts on the detectors by applying a temporal gating on the arrival time of the photons at the cost of some detection efficiency (see Figure \ref{fig_photonStatistic}a and b).   
By gating the photon arrival times with a $ \Delta t_\text{gate}=15\,\text{ns} $ time window, we reduce the quantum bit error rate (QBER) to $ e=m/n=8.3\,\% $ (see Figure \ref{fig_photonStatistic}c ). 

In order to end up with the same key string on both sides, Alice and Bob perform a key reconciliation via an authenticated classical channel, where they may leak information about the reconciled key to the public. Using privacy amplification \cite{bennett1988privacy}, the reconciled key with length $ n_r $ is shortened by a universal hash function to a final secret key length of $ n_\text{sec}=n_r-d $ by each communication party, which reduces the information available to Eve. 
An adequate hash function for this purpose can be obtained by computing $ n_\text{sec} $ publicly chosen independent random subset parities of the reconciled key and keeping their values secret \cite{bennett1992experimental}.
The number of bits $ d $ by which the key has to be shortened has to be determined concerning the maximal knowledge Eve has about the reconciled key. 

We are able to give an upper bound on Eve's knowledge without applying any restriction on the attack itself, however, requiring the transmission of true single photons, no a-priori information about the measurement bases, and fully characterized detectors on Bob's side. Further, we have to consider that the presented scheme involves an uncharacterized source on Alice side. We follow the security proof of an arbitrary uncharacterized source by Koashi and Preskill \cite{PhysRevLett.90.057902} to determine the maximal knowledge which Eve may has about the reconciled key. The proof is originally based on the BB84 protocol but due to a similar measure-and-estimate scheme of BB84 and BBM92, the security proof is valid for the presented system as well \cite{Yin2020} and holds valid even under general coherent attacks \cite{PhysRevLett.88.047902}. 
Assuming all errors of the sifted key to be caused by Eve, the length $ n_{\text{sec}} $ to which the secret key has to be shortened in order to be proven secure is given by \cite{Yin2020}:
\begin{equation}
	n_{\text{sec}} = \sum_{i \in \{z,y\}}   n_{\text{sec},i} 
	\label{Eq_secretKeyLengthBinaryEntropy}
\end{equation}  
where 
\begin{equation}
	n_{\text{sec},{z/y}}= n_{z/y}\cdot\left[1-f_r\cdot H(e_{z/y})-H(e_{y/z})\right]
	\label{Eq_AsymptoicSecretKeyrate}
\end{equation}
is the lower bound on the asymptotic secret key length for the number of sifted key bits $ n_i $ measured in the $ \sigma_z/\sigma_y $ basis respectively with $ H(q)=-q\log_2(q)-(1-q)\log_2(1-q) $ being the binary entropy function. The QBER we obtain for the sifted key bits in the $ \sigma_z/\sigma_y $ basis is $ e_{z}=7.86\,\% $ and  $ e_{y}= 9.12\,\%$. Due to timing issues in the rotated basis, $ e_y $ and $ e_z $ mismatch (see Atom state detection). 
An ideal reconciliation protocol would reveal a fraction $ H(e) $ of the sifted key, while real protocols reveal a fraction $ f_r\cdot H(e) $ with $ f_r\geq1 $ being the reconciliation inefficiency \cite{5205475}. It is convenient to introduce here the secret key rate 
\begin{equation}
r_{\text{sec}}=\frac{n_{\text{sec}}}{n},
\label{Eq_AsymptoticSecretKeyRate}
\end{equation}
which is normalised to the number of transmitted sifted key bits $ n $.

\subsection{Single photon proof}
Alice and Bob can validate the security of the key distribution via the measured QBER and the temporal $ g^{(2)}(\tau) $ correlation function of the ion as a photon source. The security proof of the presented key distribution protocol requires the transmission of true single photons in order to be applicable. Before executing the key distribution protocol, Alice has to verify the generation of single photons from the source, for example by using a Hanbury Brown and Twiss (HBT) setup. After that, the source remains under her control. Since Eve as an advanced attacker could imitate a single photon source, a similar measurement on Bob's side during the key distribution does not provide relevant information for the security of the key.

We conduct the single photon proof by measuring  the temporal second-order correlation function $ g^{(2)}(\tau) $ of the photon arrival times on a HBT setup. We obtain a value of $  g^{(2)}(0)=0.00(5) $ for the cross-correlation of the photon detection times  after correcting for background noise of the detectors according to \cite{Keller2004}, see Figure \ref{fig_photonStatistic}d and e.

\subsection{ Security measure against side channel attacks}
In general, entanglement-based QKD is source independent which has to be considered in the security proof of the applied key distribution protocol. For the presented entanglement-based QKD, we use the security proof given in \cite{PhysRevLett.90.057902} for an uncharacterised source, which requires characterised detection setups. Therefore, the presented protocol is device-dependent, i.e., we prove security for this specific setup under ideal detection conditions.
However, a real setup is vulnerable to attacks via side channels as well. We will show that we can mitigate the known detection side-channel attacks \cite{Yin2020} by taking into account the way of measurement.  
Specifically, we consider three vulnerabilities: (1) Beam splitting \cite{PhysRevA.84.062308}, (2) Efficiency mismatch between different paths \cite{PhysRevA.74.022313}, and (3) the detector dead time \cite{Lydersen2010}.

(1) The attack describes gaining control over the measurement basis on Bob's side by forcing a click on a specific detector (pair). This can be achieved by sending photons of different wavelengths, exploiting the spectral response of the setup. This attack can be mitigated by spectral filtering of the incoming photons, which we do with a $ 10 $\,nm spectral bandpass filter.

(2) Due to a difference in detection or path efficiency of the photon detection setup used, Eve may also get partial control over which detector clicks in further degrees of freedom and bias the outcome of the measurement. As a countermeasure, we implemented a series of filters on Bob's side.  Due to the photon guiding fiber being single mode, we restrict the spatial mode degree of freedom. Using 10\,nm spectral filters in front of each detector narrows the frequency degree of freedom. We restrict the time degree of freedom by applying a 15\,ns wide filter to the photon arrival time on Bob's side. We achieve equal detection efficiencies of $ H $- and $ V $- arm with an accuracy of more than 98\,\%.

(3) After the detection of a photon detectors usually exhibit a dead time in which they are blind for further photons. If there is a click followed by another click within the dead time of the detectors, it is clear to an attacker that two different detectors were involved which may reveals additional information. For example an attacker could force a detector to click in advance to a real photon detection event. We can counteract this attack by discarding runs where we noticed two clicks within one repetition period of the experimental sequence. We do this without losing QKD rate, since the probability of detecting the single photon coming from Alice in normal operation of the experiment is $P_\text{det}\approx 2.6\times 10^{-3} $ per shot while the probability of a noise count is smaller than~$ 10^{-5} $ within the dead time of the detectors  which is $ <20 $\,ns.  Assuming the transmission of true single photons, the probability of a noise click alongside with a photon detection is in the order of magnitude of $\sim  10^{-8} $.  In the presence of an attacker attempting to blind the detectors, the fraction of discarded runs would naturally be higher and the protocol would decrease in rate.

\subsection{ Privacy amplification }
Privacy amplification takes all information leakage into account by reducing the reconciled key to a length of $ n_{\text{sec}} $, which is the maximum length for the resulting secret key to be distributed provable secure (see equation (\ref{Eq_secretKeyLengthBinaryEntropy})). However,  $ n_{\text{sec}} $ states the asymptotic secret key length for large $ n $. For a finite key length, we obtain a non-vanishing finite secret key rate for a failure probability of $ \epsilon \approx 3\,\% $ as follows: Due to statistical fluctuations, the measured error rate $ e_{z/y} $ may differ from the underlying error rate $ e'_{z/y} $ obtained in the asymptotic case of large $ n $. An upper bound $ e^\text{max}_{z/y} $ of the underlying error rate $ e'_{z/y} $ can be given using the Serfling inequality \cite{serfling1974probability} according to \cite{Yin2020, curty2014finite}:
\begin{equation}
	e'_{z/y}\leq e^\text{max}_{z/y}=e_{z/y}+\sqrt{\frac{\left(n_{z/y}+1\right)\log\left(1/\epsilon_\text{sec}\right)}{2n_{z/y}\left(n_{z/y}+n_{y/z}\right)}}
\end{equation}
where $ n_{z/y} $ is the number of measured sifted key bits in $ \sigma_{z/y,\text{photon}}\otimes \sigma_{z/y,\text{atom}}$ basis.
The failure probability $ \epsilon=\epsilon_\text{sec} + \epsilon _\text{ec} $ ($ \epsilon_\text{sec} $: secure transmission of the key fails, $ \epsilon_\text{ec} $: error correction fails) can be used according to \cite{Yin2020} in order to derive a finite secret key length for which the security proof still applies. 
Using the upper bound of the bit error rate $ e $, we can compute the finite secret key rate according to \cite{tomamichel2012tight,Yin2020} as:
\begin{equation}
	r^\text{finite}_{\text{sec}} = \sum_{i \in \{z,y\}}   r^\text{finite}_{\text{sec},i} 
\end{equation}  
with
\begin{equation}
	r^\text{finite}_{\text{sec},z/y}= r_{z/y}\cdot\left[1-f_r\cdot H(e_{z/y})-H(e^\text{max}_{y/z})\right]-\log\left(\frac{2}{\epsilon_\text{ec}\epsilon^2_\text{sec}}\right).
\end{equation}

To form a secure key in practice, we apply the symmetric blind information reconciliation protocol \cite{PhysRevApplied.8.044017}, which is based on low density parity checks (LDPC) on the sifted key obtained on the sides A and B.  We reach a reconciliation inefficiency of $ f_r=1.16 $ and subsequently $ r_\text{sec} =0.096$\,bits according to equation~(\ref{Eq_secretKeyLengthBinaryEntropy}). With a sifted key bit rate of 6\,Hz, this corresponds to an asymptotic secret key bit rate of 0.6\,Hz. Finally, privacy amplification has to take all information leakage into account by reducing the reconciled key to a length of $ n_\text{key}$ with $n_\text{key}\leq n_\text{sec} $. In particular, we have to shorten the secret key to the length \begin{equation}
 n_\text{key} \leq \min \{n_\text{sec},n_\text{rand}\},
\end{equation} 
which is the smaller of the two maximal allowed key lengths to obtain: i) a certified random generated key with length~$ n_\text{rand}=n\cdot H_{{\text{min},\delta,k}}^{\geq} $ and ii) a provably secure transmitted key with length~$ n_\text{sec}^{(\text{finite})}=n\cdot r_\text{sec}^{(\text{finite})} $.

\subsection{Characteristic of the memory qubit}
In the following, we investigate the influence of the memory qubit on the key distribution protocol, i.e on the achievable QBER, the resulting key rates and its applicability in a repeater like setup. A crucial property of quantum memories is the coherence time of the memory quantifying the dephasing of the stored quantum information. This quantity dictates the achievable storage time, after which the information can be retrieved from the memory with an acceptable error. 
In purely photon-based QKD methods, where the quantum information can be encoded into different degrees of freedom of one or more photons, the coherence time usually exceeds the storage time of quantum information, since the latter is bound to the lifetime limit of the photons imposed by absorption, which ultimately destroys the information. This directly sets a limit to the distance that can be bridged for key distribution. 
However, the opposite is true for matter qubits, which are very sensitive to environmental influences, but have availabilities that extend beyond the signal propagation path of the key distribution and can even be used for the entire key transmission. Here the limit on the distance is set by the coherence time of the memory qubit.
 
 In the presented scheme, a photon is directly transmitted to the receiver while the memory resides on the sender's side. 
In this case, the dephasing of the memory qubit directly affects the QBER as an erroneously retrieved quantum state likely causes a bit error in the sifted key on the sender's side. 
The corresponding phase error of the qubit state becomes most dominant for measurements in $ \sigma_y $ basis, where timing uncertainties ($\leq 7.8(2) $\,ns) of the experiment comes into play due to the Larmor precession of the spin superposition state in laboratory frame ($5.477(1)\, $MHz). Furthermore, this precession frequency exhibits a noise contribution due to magnetic field fluctuations (0.9(1)\,mG peak-to-peak) at the position of the memory qubit. For increasing storage times $ T_s $, we found that the QBER becomes mainly dominated by this magnetic field noise.

\begin{figure}
   	\centering
   	\includegraphics[width=0.95\columnwidth]{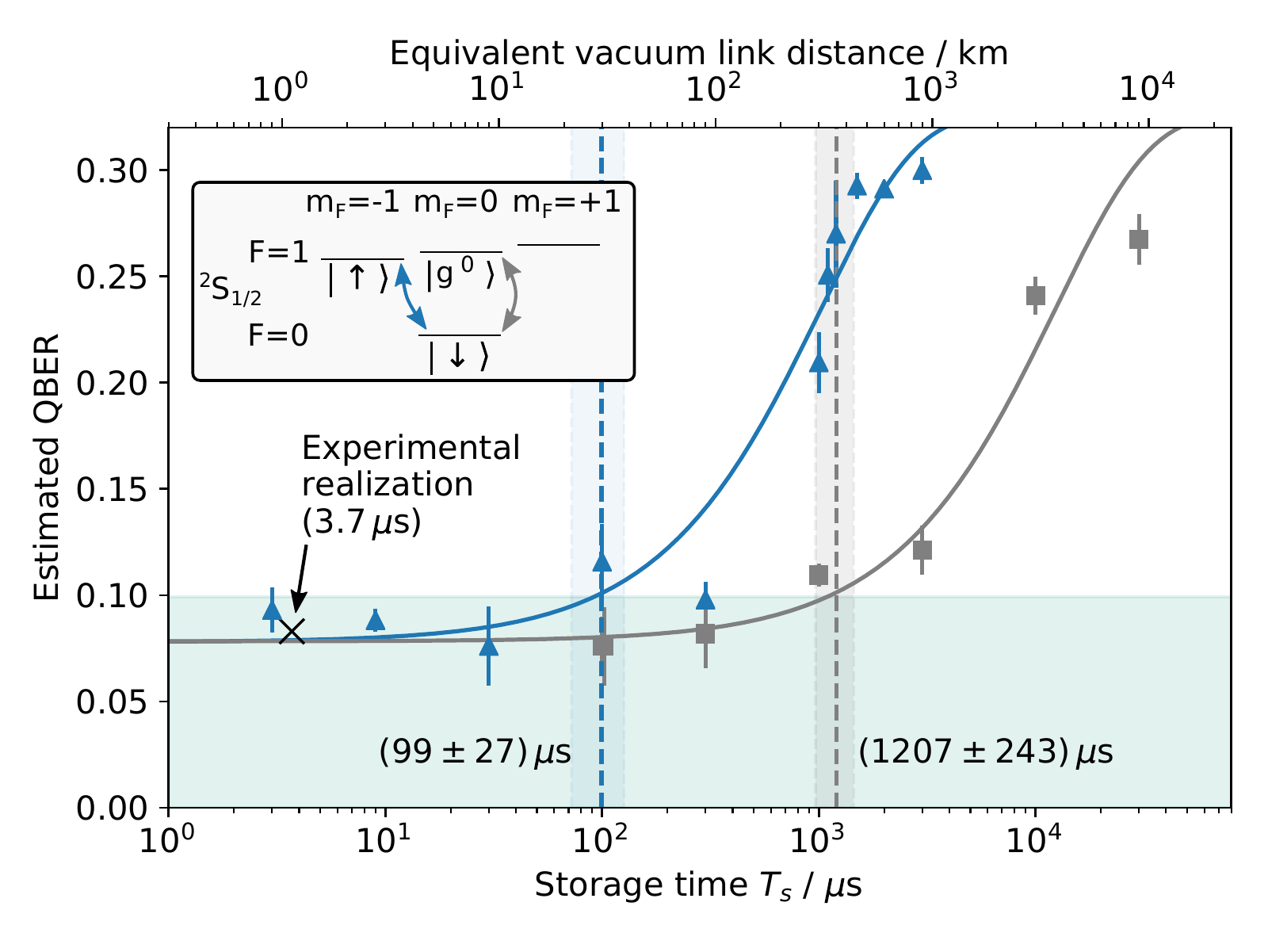}
   	\caption{{\bf Calculated QBER vs. storage time of quantum information.} The data points originate from a coherence time measurements of the respective atomic qubit representing the influence of the magnetic field noise and are converted to an expected QBER taking into account the experimental imperfections (see Table \ref{Table_ContrastNEgativrInfluence}). Triangles: $\ket{ \downarrow/\uparrow} $-qubit, squares: $ \ket{\downarrow/g^0} $-qubit. The equivalent vacuum link distance is given for the storage time $ T_s $, where we excluded potential absorption and conversion inefficiencies of the photon as travelling qubit from the calculation. The solid lines were derived from fits according to $ \exp\left(-T_s/\tau\right)$ to the coherence time data of the qubits. The green shaded area depicts the region of a non-vanishing secret key rate according to equations~(\ref{Eq_secretKeyLengthBinaryEntropy})-(\ref{Eq_AsymptoticSecretKeyRate}) (QBER$<9.92\,\% $). Vertical dashed lines represent the corresponding maximal storage time of quantum information in the memory qubit with the standard error given as shaded area. Inset: Relevant energy levels of the trapped ion. The investigated qubit transitions are labelled with the corresponding coloured arrow. 
   		} 
   	\label{fig_QBER_storageTime}
   \end{figure}

We put together the major contribution to the correlation contrast of the atom-photon state in Table \ref{Table_ContrastNEgativrInfluence}. Figure \ref{fig_QBER_storageTime} shows the expected QBER of the presented protocol as a function of the storage time derived from the correlation contrast. Using the clock transition $ \ket{\downarrow} \leftrightarrow \ket{g^0}\equiv \ket{F=1,m_F=0}$ as a memory qubit, we can achieve non-zero asymptotic secret key rates for a maximum of $ (1.2\pm 0.3) $\,ms according to equation (\ref{Eq_AsymptoicSecretKeyrate}). Theoretically, this would be sufficient for a link distance of around $400 $\,km. 
\begin{table*}[!htb]
	\centering
	\begin{tabular}{l| c | c  }

		Source &$ \bar{\hat{\sigma}}_y\otimes  \bar{\hat{\sigma}}_y $&$  \hat{\sigma}_z\otimes \hat{\sigma}_z $ \\
		\hline
		Atomic qubit manipulation	& $ \leq1.9 $ &$\leq1.9  $ \\
		\hline
		Atomic state discrimination	&$  3.5\pm1.2 $ &$  3.5\pm1.2 $ \\
		\hline
		Timing of atomic readout 	&$ \leq7.0\pm0.1 $&- \\	
		\hline
		Magnetic field noise (@$ T_s=3.7\,\mu$s) & $ 0.7\pm0.1$  & - \\
		\hline
		Atomic excitation &$  4.2\pm2.8 $ & $  4.2\pm2.8 $\\ 
		\hline
		Basis selection on Bob's side (photon polarization)& $ \ll1.0 $  &$  \ll1.0 $ \\ 
		\hline						
		False detection events on Bob's side & $ 3.3\pm0.1 $  &$  3.3\pm0.1 $ \\ 
		(dark counts)&&\\

	\end{tabular}
	\caption{Sources of error of the entangled state measurements with associated values of the correlation contrast reduction in \% broken down by the measurement bases. }
	\label{Table_ContrastNEgativrInfluence}
\end{table*}

\section*{Discussion}

We have demonstrated the realisation of QKD between two remote parties including an entangled memory qubit on one side, which favours long-distance key exchange, especially with respect to quantum repeaters. 
According to the laws of quantum mechanics, the secure distribution of the final key is guaranteed by the transmission of true single photons, which we prove with $ g^2(0)=0.00(5) $. 

Considering the fundamental non-locality of our two-qubit entangled state for the generation of the final secret key, we have shown that certifiable randomness of the derived key can be ensured with a high confidence level $ \delta<0.01 $ by performing a finite set of Bell test measurements $O (10^4) $. As a lower bound on the generated randomness we calculated $  H_{\text{min}}^{\geq}=0.15(4)  $ bits per sifted key bit in the asymptotic limit. This kind of provable randomness is an outstanding property of entangled quantum systems violating the Bell inequalities and is impossible to obtain in classical information theory.  
We measure a Bell-violation of $ g_\text{meas}=2.33(6)$, which is consistent with the measured entangled state fidelity of $ 90.1(17)\,\% $. 

For a sifted key with a length of~$ n=3080 $, we have achieved a quantum bit error rate (QBER) of $ 8.3\,\%$, which is expectable from the experimental imperfections and mainly limited by the state detection of the stationary qubit.
The measured sifted key rate of 6\,Hz depends on the experimental repetition rate ($ \sim 20\, $kHz) and the channel loss. We achieve $ 1.29(3)\times 10^{-3} $ sifted key bits per channel use (per atomic initialisation) which is more than three orders of magnitude higher than the sifted key rate reported for a non-distant quantum communication including an entangled memory qubit with a QBER of $ \sim 11\,\%$ \cite{Bhaskar2020}. We calculated an asymptotic secret key rate of $ r_\text{sec} =0.096$~bits per sifted key bit which transforms to approximately $ 1.3\times 10^{-4} $ secret key bits per channel use. This is comparable to the secret key rate reported for a non-distant QKD involving two entangled memory qubits with a QBER of $ \sim 8\% $ \cite{PhysRevLett.126.230506}. 
We would like to emphasize that the presented secret key rate is fast enough to update a secret key several times per hour, which is sufficient for bipartite communication. Depending on the use case, this could even be sufficient to implement perfect forward secrecy.

On the one hand, the use of memory qubits allows arbitrarily long communication distances within the framework of an ideal quantum repeater, but on the other hand requires comparatively long readout and preparation times, which in our case lead to a sifted key rate about two orders of magnitude lower compared to QKD implementations with entangled photon pairs \cite{Erven:08,doi:10.1063/1.2348775,Zhong_2015}. There, the best achieved error rates are about a factor of 2 lower ($ \gtrsim 4.5\,\%$) \cite{Yin2020,Ursin2007,Peloso_2009,Erven:08} than the QBER presented here.

The security of the presented system in terms of key randomness and tapproofing is based on the violation of Bell's inequality, photon correlation measurement and error rate estimation (see Figure \ref{fig_intro}b). It is also possible to combine the measurement blocks into a single sequence, which requires a small modification to the measurement bases of the presented protocol so that they maximally violate the CHSH Bell inequality. Alice and Bob can then perform permanent Bell tests on the quantum system to ensure random generation and secure distribution of the key, as proposed in \cite{PhysRevLett.113.140501,Masanes2011}. 

We showed that entanglement-based QKD can combine the crucial properties of secret key distribution, namely  confidentiality and integrity with the required properties of key generation, namely certifiable randomness.
The presented methods can in principle be extended to any two-qubit entangled state and are particularly applicable in the context of entanglement-based quantum repeaters that benefit from memory-enhanced quantum communication. 
For this purpose, the presented system is advantageous due to the intrinsic fiber coupling of transmitted qubits, which allows for efficient distribution of information to the communication parties.

We thank E. Kiktenko for the useful discussion and for providing the LDPC error codes to us. We thank D. Bruss and H. Kampermann for insightful discussions. This work has been funded by the Alexander-von-Humboldt Stiftung, DFG (SFB/TR 185 project A2), BMBF (FaResQ and Q.Link.X), and the Deutsche Forschungsgemeinschaft (DFG, German Research Foundation) under Germany's Excellence Strategy – Cluster of Excellence Matter and Light for Quantum Computing (ML4Q) EXC 2004/1 – 390534769.
\newline

%

\end{document}